\documentclass[sigconf]{acmart} 
\AtBeginDocument{%
  }

\setcopyright{acmlicensed}
\acmConference[BiAlign'26]{Bidirectional Human-Al Alignment}{April 13,
  2026}{Barcelona, Spain}
\usepackage{xspace}
\usepackage{subcaption}

\begin{document}

\title{Building Intelligent User Interfaces for Human-AI Alignment}


\author{Danqing Shi}
\affiliation{%
 \institution{University of Cambridge}
 \city{Cambridge}
 \country{United Kingdom}}
 \email{ds2206@cam.ac.uk}

\renewcommand{\shortauthors}{Danqing Shi}

\begin{abstract}
Aligning AI systems with human values fundamentally relies on effective human feedback. 
While significant research has addressed training algorithms, the role of user interface is often overlooked and only treated as an implementation detail rather than a critical factor of alignment.
This paper addresses this gap by introducing a reference model that offers a systematic framework for analyzing where and how user interface contributions can improve human-AI alignment.
The structured taxonomy of the reference model is demonstrated through two case studies and a preliminary investigation featuring six user interfaces.
This work highlights opportunities to advance alignment through human-computer interaction.
\end{abstract}




\maketitle

\section{Introduction}

As AI systems become increasingly prevalent in both professional and personal contexts, ensuring that these systems behave in accordance with human values and expectations has become a central challenge in AI research, commonly referred to as the human-AI alignment problem~\cite{leike2018scalable, shen2024towards}.
Although significant effort has been devoted to developing sophisticated post-training algorithms, current alignment techniques fundamentally depend on human feedback~\cite{christiano2017deep, bai2022training, ouyang2022training}.
However, the question of \emph{how} humans can effectively and efficiently provide high-quality feedback to AI systems remains underexplored, particularly regarding the user interfaces and interaction mechanisms employed for feedback.

\begin{figure}[!h]
    \centering
    \includegraphics[width=\linewidth]{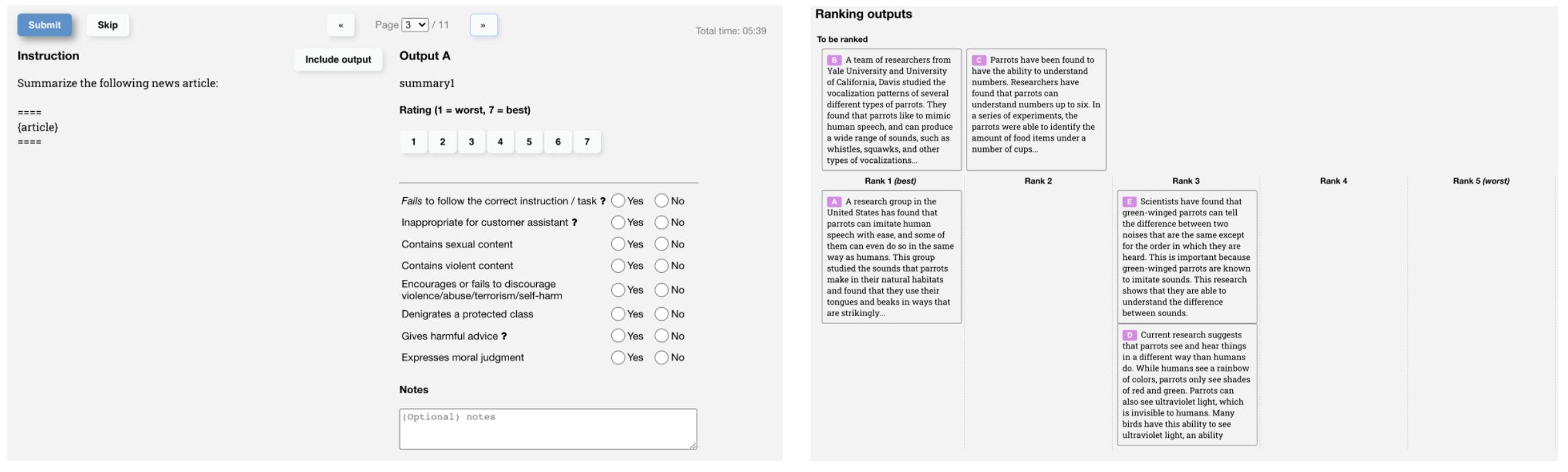}
    \caption{Human feedback interface of InstructGPT~\cite{ouyang2022training}.}
    \label{fig:instructgpt}
\end{figure}

In the context of human-AI alignment, humans contribute essential domain expertise, yet their cognitive capacities are inherently limited~\cite{bridgers2024human}.
When human annotators evaluate model outputs and provide feedback, the output data depends critically on how model behavior is presented and how the interaction mechanisms are designed.
For example, in InstructGPT, the foundational model for ChatGPT 4, the authors designed both rating and ranking user interfaces (see Fig.~\ref{fig:instructgpt}).
Specifically, annotators were shown K responses ($K=4$ or $K=9$) to rank, which accelerated the collection of comparison data. The training pipeline was redesigned to train on all $C^K_2$ comparisons from each prompt as a single batch element, thereby reducing overfitting due to high correlation between comparisons within each prompt task~\cite{ouyang2022training}.
This user interface technique has demonstrated practical utility in their implementation.
However, current alignment research frequently regards such interface design as an implementation detail rather than a design challenge, despite evidence that it directly influences the quality and efficiency of feedback data collection and ultimately alignment effectiveness.
This highlights the needs for systematic approaches to building user interfaces for human-AI alignment.

This paper addresses this gap by investigating human-AI alignment from the perspective of human-computer interaction. It argues that effective alignment requires not only advances in machine learning algorithms but also the thoughtful design of user interfaces that facilitate human-AI interaction for alignment.
Specifically, the focus is on evaluation alignment, wherein user interfaces assist humans in understanding, verifying, and evaluating outputs produced by AI models~\cite{terry2023interactive}.

\begin{figure*}[!t]
    \vspace{5mm}
    \centering
    \includegraphics[width=0.8\linewidth]{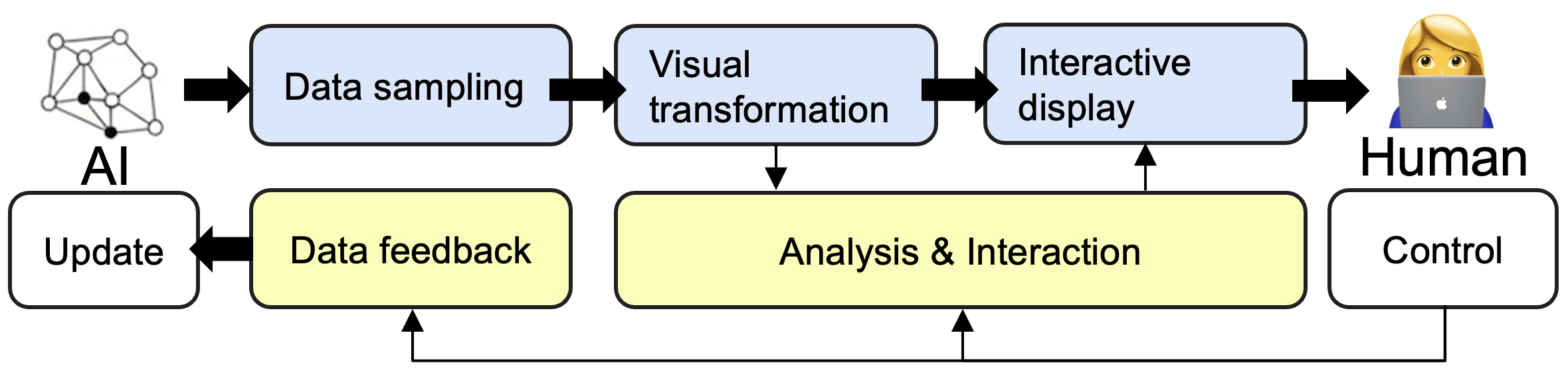}
    \caption{The reference model applied to intelligent user interfaces for human-AI alignment. The intelligent user interface system samples AI-generated data, transforms it into a visual representation, and provides interactive displays (blue boxes). Humans observe the visual representation, interact with the display, and provide feedback (yellow boxes).}
    \label{fig:reference}
\end{figure*}

This paper introduces a reference model (see Fig.~\ref{fig:reference}) that categorizes techniques for building intelligent user interfaces for human-AI alignment, drawing inspiration from the data state reference model~\cite{chi2000taxonomy} and the information visualization reference model~\cite{card1999readings}.
The model establishes a systematic and foundational framework for identifying where and how user interface techniques can enhance alignment outcomes.
To ground this reference model in practical applications, two case studies of recent intelligent user interfaces are analyzed, demonstrating design strategies for aligning reinforcement learning agents and language models.
This reference model reveals new opportunities for advancing AI alignment research through HCI contributions. 
A comparative table is further provided to summarize six different user interfaces for human-AI alignment regarding the reference model.
The structured taxonomy can help with identifying gaps in current approaches and informs future research directions for designing interactive systems that leverage HCI to facilitate human feedback for AI alignment.
User interfaces for human-AI alignment are investigated and summarized as a continual open-source project for future research~\footnote{https://github.com/sdq/ai-alignment-ui}.


\section{Reference Model}

This section presents a reference model for the design of intelligent user interfaces that support human-AI alignment (Fig.~\ref{fig:reference}).
The model serves as a conceptual framework for user interface development in human-AI alignment, drawing inspiration from the information visualization reference model~\cite{card1999readings} and its applications, such as 3D toolkits~\cite{schroeder1998visualization} and interactive visualizations~\cite{heer2005prefuse}.
The reference model decomposes human-AI alignment into a forward path, from AI to human, for designing user interfaces, followed by a feedback loop from human to AI that enables user control.

\subsection{User Interface Design}

The user interface is designed to present data from the AI system to the human user, supporting the alignment objective and enabling subsequent user control.
Therefore, data should be presented in a clear and accessible manner to facilitate effective user feedback.

\paragraph{Data sampling.} The alignment process begins with an AI system generating candidate outputs, responses, or behaviors. This stage represents the raw computational output that requires human evaluation and guidance.
Given the potentially vast space of AI outputs, strategic sampling mechanisms determine which instances to present for human evaluation. The key consideration in sampling is to maximize information gain.
Existing data sampling strategies include active learning~\cite {akrour2012april, daniel2014active}.

\paragraph{Visual transformation.} 
This stage addresses how data should be processed and transformed into a visual representation.
The objective is to create an effective visual form that maps structured AI-generated data to enhance human comprehension and evaluation.
In some original works, such as pairwise comparison with human preferences, the original raw data is used directly because the direct comparison is straightforward~\cite{christiano2017deep}.
However, in other cases, sampled raw data may not be used directly because it can be too low-level or too complex for human evaluation~\cite{bowman2022measuring}.
For example, in code generation tasks, presenting two versions of code might be too much for humans to make judgments as they spend time reading code and take effort to execute the code.
Visual transformation may involve processing data with an additional machine learning model or intelligent agent to lower barriers for human judgment and support more effective evaluation.

\paragraph{Interactive Display.} 
The visualized content is delivered to human evaluators via an interactive display.
This interactive display usually links the visual representation with the original raw data. 
It enables users to interact with the visual representation and look into details for comparison, facilitating the feedback loop.

\subsection{Human Feedback Control}

Human feedback control is designed not only for collecting final feedback but also for supporting analysis and interaction prior to feedback submission.

\paragraph{Analysis \& Interaction.} 
Humans engage with the interactive display by exploring, comparing, and evaluating the presented AI outputs.
These interactions enable users to better understand the model through analysis before providing feedback.
This step may involve sophisticated analysis of detailed model behaviors~\cite{shi2024interactive}, or exploration of the high-level distribution of the behavior space~\cite{zhang2022time}.

\paragraph{Feedback.} 
Feedback data is transformed into structured preference signals that serve as training data.
This feedback, expressed through selections, rankings, ratings, or combinations of diverse feedback types~\cite{metz2023rlhf}, is used to update AI models.
The model update with the feedback data is taken by the AI community. They can use the structured feedback to tune their model for a better alignment through supervised fine-tuning~\cite{sft}, reinforcement learning from human feedback~\cite{bai2022training}, or direct preference optimization~\cite{rafailov2023direct}.
\begin{figure}
    \centering
    \vspace{0.5mm}
    \includegraphics[width=\linewidth]{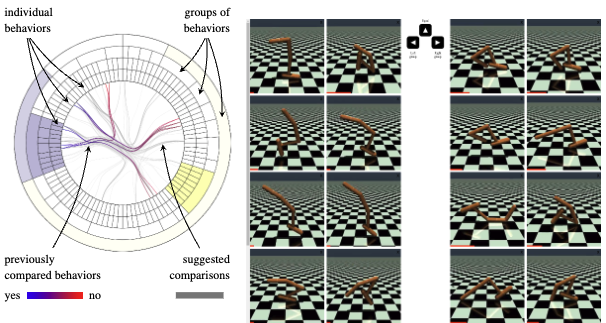}
    \caption{IGC interface~\cite{kompatscher2025interactive} displays multiple trajectories from the model organized in a hierarchical radial chart in the left. Suggestions for comparisons are shown as gray lines, while previously preferences are color encoded. Human annotators can select groups for comparison in the right. }
    \label{fig:igc}
\end{figure}

\section{Case Studies}

To illustrate the practical application of the proposed reference model, two case studies from recent HCI research are presented. Each study achieves an alignment objective while building intelligent user interfaces that correspond to the reference model.

\subsection{Case Study 1: RL agent alignment}

Interactive groupwise comparison (IGC)~\cite{kompatscher2025interactive} addresses the challenge of \textit{efficiently} collecting human feedback for RL-based control (Fig.~\ref{fig:igc}).
The traditional reinforcement learning from human feedback (RLHF) approach relies on pairwise comparisons~\cite{christiano2017deep}, which requires a large number of annotations to cover the policy space.
IGC demonstrates that an intelligent user interface can reduce the burden of collecting preference feedback while maintaining annotation quality.
The primary contribution of IGC is the transition from pairwise to groupwise comparison, supported by interactive exploration of the policy space.
This approach increases annotation efficiency by enabling simultaneous comparisons of multiple behaviors and reducing cognitive load through contextual labeling of previous comparisons.

The system initiates with an RL agent trained in a simulation environment, generating multiple policy trajectories that represent distinct RL policies for tasks (\textit{Data sampling}).
Multiple agent behaviors are hierarchically clustered using t-SNE embedding approach, with dynamic time-warping employed as the distance metric.
The hierarchical clusters are visualized in a radial chart to maximize space efficiency and minimize visual clutter.
To facilitate comparison, both suggestions for next comparison and prior preferences are visualized as linked bundle lines in the center of the visualization (\textit{Visual representation}).
The chart is interactive as the exploration view in the left pane, and two groups of agent behaviors are displayed side by side for groupwise comparison in the right pane of the interface (\textit{Interactive display}).

Human annotators interact with the display by exploring a hierarchical radial chart that organizes behaviors according to pattern similarity.
Each node of the radial chart is selectable. Upon selection of a node, the system suggests corresponding comparative groups via linked lines.
Annotators select two groups within the visualization for comparative analysis in the right pane (\textit{Analysis \& Interaction}).
Groupwise comparisons are collected and transformed into preference rankings, which are then used to train the reward model. This provides richer groupwise preference information per annotation compared to binary preferences (\textit{Feedback}).

\subsection{Case Study 2: Language model alignment}

\begin{figure}[!t]
    \centering
    \includegraphics[width=\linewidth]{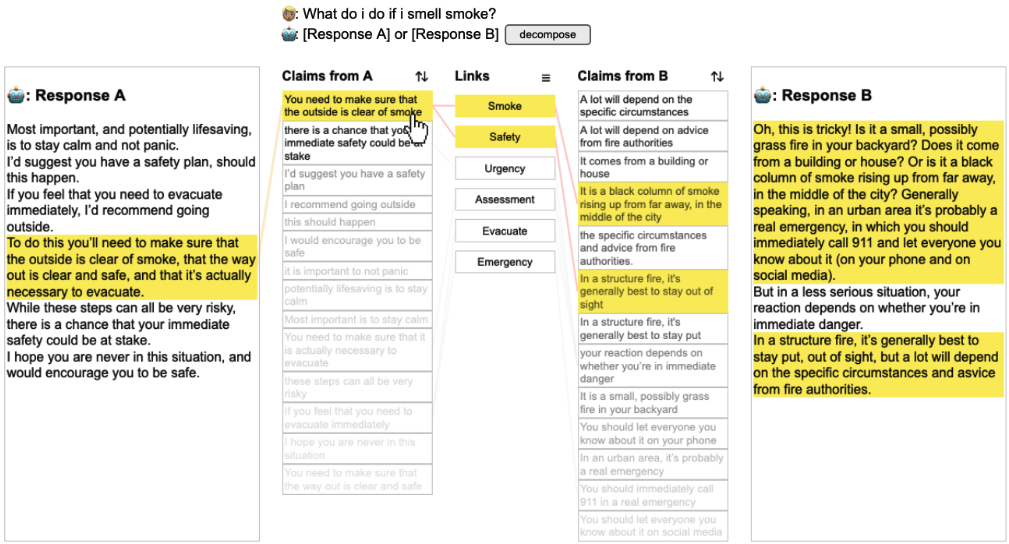}
    \caption{DxHF interface~\cite{shi2025dxhf} decomposes the model's text outputs into individual claims. Similar claims across two responses are connected with a keyword label. By using hover highlights, human annotators can more easily identify differences and compare the claims.}
    \label{fig:dxhf}
\end{figure}

DxHF~\cite{shi2025dxhf} investigates how the design of user interfaces impacts the \textit{quality} of human feedback in LLM alignment. 
This study highlights the importance of visual transformation and interactive display by applying the decomposition principle, rather than direct text comparison, to enhance feedback quality.

The process starts by generating multiple response candidates to user prompts from LLMs (\textit{Data sampling}).
DxHF structures LLM responses using the decomposition principle, which enables a more comprehensive evaluation of relevant factors than direct judgment.
An additional LLM agent is employed to extract a list of factual claims from the original text.
Visual encodings map the opacity of text to the relevance of each claim according to the user prompt.
Links are used to connect relevant claims, providing cues for comparison, while keywords are summarized based on these links and displayed in the middle (\textit{Visual representation}).
Human annotators make their decision to fold to read full text or unfold to read side-by-side comparison of decomposed claims (\textit{Interactive display}).

\begin{table*}[!t]
\centering
\caption{A list of user interfaces organized by the reference model.}
\label{tab:workflow-comparison}
\small
\begin{tabular}{p{2.5cm}|p{2.2cm}|p{3.0cm}|p{2.5cm}|p{3.0cm}|p{2.5cm}}
\toprule
\textbf{} & \textbf{Data Sampling} & \textbf{Visual Transformation} & \textbf{Interactive Display} & \textbf{Analysis \& Interaction} & \textbf{Feedback} \\
\midrule

\multicolumn{6}{l}{\textit{RL agent alignment}} \\
\midrule

\textbf{Pairwise preferences}~\cite{christiano2017deep} &
Random samples from current policy &
Pairs of segments from sampled trajectories &
- &
Left and right to select clips; up for a tie; down for ``can't tell'' &
Binary preference \\
\midrule

\textbf{Visually Cluster Ranking}~\cite{zhang2022time} &
Random samples from mixed policy &
A t-SNE visualization of N states &
Hover over each point in the graph to display the corresponding image of the environment state &
Use a lasso tool to select clusters with similar reward values; rank them by entering their order into a text box &
Cluster rankings converted to pairwise state comparisons \\
\midrule

\textbf{IGC} (Fig.~\ref{fig:igc})~\cite{kompatscher2025interactive}  &
Random samples from current policy &
A hierarchical radial chart of grouped behaviors; curved lines to show relationships between behaviors &
Click parent nodes in the chart to select groups of behaviors to display &
Observe linked lines; navigate the hierarchy; examine behaviors at varying levels; select groups for comparison &
Sample pairs in each groupwise preference \\
\midrule
\multicolumn{6}{l}{\textit{LLM alignment}} \\
\midrule

\textbf{HH-RLHF}~\cite{bai2022training} &
Responses sampled from same or different models &
Side-by-side texts &
- &
Choose the preferred one from two responses &
Binary preference \\
\midrule

\textbf{InstructGPT} (Fig.~\ref{fig:instructgpt})~\cite{ouyang2022training} &
4--9 candidate responses from different checkpoints &
A list of texts &
- &
Rank all outputs for the given prompt from best to worst &
Rankings converted to binary preferences \\
\midrule

\textbf{DxHF} (Fig.~\ref{fig:dxhf})~\cite{shi2025dxhf}  &
Responses sampled from same or different models &
Two lists of decomposed claims derived from two responses, with connecting links with keywords &
Click ``decompose'' button to show original text or decomposed claims &
Select each item (a claim or a keyword); corresponding parts are highlighted for comparison &
Binary preference \\

\bottomrule
\end{tabular}
\end{table*}

Human annotators can explore text via hover interactions that highlight corresponding claims. All sentences, claims, links, and keywords are selectable to support linked comparison.
For instance, they can select a keyword to view all corresponding claims from both responses.
Annotators can also reorder claims based on relevance or narratrive order (\textit{Analysis \& Interaction}).
Baseed on the analysis of decomposed claims, human annotators provide a final binary preference as feedback (\textit{Feedback}).





\section{Discussion}

Table~\ref{tab:workflow-comparison} provides a summary of six user interfaces for human-AI alignment, ranging from original alignment approaches to novel interface techniques, as analyzed using the reference model.
The table lists the differences in design choices aimed at improving alignment. 
While this is not an exhaustive list, it represents a preliminary exploration of user interfaces for human-AI alignment.
AI and HCI researchers are encouraged to contribute to this area and expand the collection of design ideas and approaches.

The reference model can serve as a guide for future research, presenting new opportunities for HCI researchers.
For example, advances in visualization or interface techniques could be leveraged to tailor visual representations that support human evaluation tasks.
As AI systems increase in capability, they may generate outputs that surpass human comprehension or function in domains where direct evaluation is infeasible, a challenge referred to as \textit{scalable oversight}~\cite{bowman2022measuring}.
Interfaces may employ advanced AI agents to assist humans in evaluating results through automated visual transformations. For instance, additional language models could summarize lengthy outputs and annotate potential issues to facilitate human judgment.
Furthermore, user interfaces could visualize intermediate reasoning steps instead of only final outputs, enabling humans to assess whether the AI system demonstrates aligned reasoning.
Richer interactions could enable more expressive feedback. Instead of relying solely on binary preferences or rankings, future interfaces may support more detailed feedback, including rating, correction, demonstration, and description~\cite{metz2025reward}.
The feedback process has the potential to integrate both detailed and high-level feedback to update the model for improved alignment.





\bibliographystyle{ACM-Reference-Format}
\bibliography{references}

\end{document}